\documentclass[aps,prd,preprint]{revtex4-1}
\usepackage{amsmath,amssymb,graphicx,hyperref,float}
\usepackage[small,flushleft,indent]{caption}

\preprint{}

\begin{document}

\title{Revisiting a rotating black hole shadow with astrometric observables}

\author{Zhe Chang}
\author{Qing-Hua Zhu}
\email{zhuqh@ihep.ac.cn}

\affiliation{Institute of High Energy Physics, Chinese Academy of Sciences, Beijing 100049, China}
\affiliation{University of Chinese Academy of Sciences, Beijing 100049, China}

\date{\today}

\begin{abstract}
	The first image of black hole in M87 galaxy taken by Event Horizon
	Telescope shows that directed observation of supermassive black holes would be a
	promising way to test general relativity in strong gravity field regime. In
	order to calculate shadow of rotating black holes with respect to observers
	located at finite distance, orthonormal tetrads have been introduced in previous works. However, it is noticed that different choice of tetrads does not lead to the same 
	shape of shadow for observers in near regions. In this paper, we
	alternatively use formula of astrometric observables for calculating the
	shadow of a general rotating black hole with respect to these observers. 
	For the
	sake of intuitive, we also consider Kerr-de Sitter black holes as a representative example. In this space-time, size and shape of Kerr-de Sitter black hole shadows are expressed as functions of
	distance between the black hole and observer. It is forecasted that the distortion of shadow would
	increase with distance.
\end{abstract}

\maketitle

\section{Introduction}

In strong gravity field regime, one of the most interesting predictions of
General relativity might be black holes. Besides existence of stellar-mass
black holes nearly to be confirmed by observation of gravitational wave \cite{Abbott:2016nmj}, there
is also development of directed observation of supermassive black holes in the
centre of galaxy, such as the first sketch of black hole in M87 galaxy taken by
Event Horizon Telescope (EHT) \cite{collaboration_first_2019}.  And for observation in the future besides the EHT, the $BlackHoleCam$ project also targets on making
images for Sagittarius A*, the supermassive black hole in the centre of Milky
Way \cite{goddi_blackholecam_2016}.

Due to highly bending light rays in strong gravity field, the black holes usually cast
a shadow in the view of observers.  It has been shown by the first sketch of
black hole \cite{collaboration_first_2019}. What the observers perceive  are light rays from unstable circle orbit called photon sphere or photon region.  In the sixties of last century, Synge \cite{synge_escape_1966} firstly studied the
shadow of Schwarzschild black holes. 
To date, the studies referred to shadow of a
spherical black hole are still non-trival \cite{wang_shadows_2018,tian_testing_2019,konoplya_shadow_2019}. Besides exotic matter and modified gravity \cite{zhu_shadows_2019,allahyari_magnetically_2020}, one can consider how the expansion rate of the universe
would affect size of black hole shadow \cite{perlick_black_2018,bisnovatyi-kogan_shadow_2018,tsupko_first_2019,firouzjaee_black_2019,Chang:2019vni,vagnozzi_concerns_2020}. On the other side,
Bardeen \cite{bardeen_timelike_1973} firstly shown that the spin of Kerr black holes would cause the shape of shadows
distorted. It was originally understood as frame dragging effect on the shadow
and  expected to be observed in the future \cite{falcke_viewing_1999}. 
Recent works also considered extended Kerr black holes, such as Kerr-de Sitter black holes
\cite{grenzebach_photon_2014,stuchlik_light_2018,Li:2020drn}, deformed black holes \cite{atamurotov_shadow_2013},
regular black holes \cite{li_measuring_2014,abdujabbarov_shadow_2016}, superspinars \cite{bambi_testing_2019}, accelerated Kerr black holes \cite{grenzebach_photon_2015}, Kerr black holes in the presence of extra dimensions \cite{vagnozzi_hunting_2019,banerjee_silhouette_2020} or surrounded by dark matter \cite{jusufi_black_2019} and Kerr black holes coupled to scalar hair \cite{cunha_shadows_2015,cunha_eht_2019},
perfect fluid dark matter \cite{haroon_shadow_2019}, axion field \cite{wei_intrinsic_2019,bar_looking_2019,davoudiasl_ultra_2019}, quantized bosonic fields \cite{roy_evolution_2020} or background vector field \cite{li_testing_2019}. 

In asymptotic flat space-time, calculation of rotating
black hole shadow for distant observers is simple. The formula of angular radius
of shadow is the same as that in Minkowski space-time
\cite{hioki_hidden_2008,hioki_measurement_2009,johannsen_photon_2013,atamurotov_shadow_2013,li_measuring_2014,abdujabbarov_coordinate-independent_2015,abdujabbarov_shadow_2016,ovgun_shadow_2018,wei_intrinsic_2019,kumar_shadow_2019,jusufi_black_2019,li_testing_2019}, namely,
\begin{eqnarray}
	\alpha_{\rm {Dist}} & = & \lim_{r_o \rightarrow \infty} \left( - r_o \sin
	\theta_o \left. \frac{{\rm d} \phi}{{\rm d} r} \right|_{\theta = \theta_o}
	\right) ~, \label{1}\\
	\beta_{\rm {Dist}} & = & \lim_{r_o \rightarrow \infty} \left( r_o \left.
	\frac{{\rm d} \theta}{{\rm d} r} \right|_{\theta = \theta_o} \right) ~ , \label{2}
\end{eqnarray}
where $(r_o, \theta_o)$ is the position of observer, $\frac{{\rm d}
		\phi}{{\rm d} r}$ and $\frac{{\rm d} \theta}{{\rm d} r}$ are used to describe
motion of light rays, $\alpha_{\rm {Dist}}$ and $\beta_{\rm {Dist}}$ are
angular radius of shadow in two orthonormal directions, approximately.
However, in the presence of a cosmological constant, the non-asymptotic flat
space-time does not allow observers located at spatial infinity. And in this case,
the Eqs.~(\ref{1}) and (\ref{2}) would not be
valid. To deal with this difficulty, one may introduce
orthonormal tetrads for observers located at finite distance. In a pioneer work
on Kerr black hole shadow, Bardeen \cite{bardeen_timelike_1973} calculated the shape of shadow via
introducing orthonormal tetrads with respect to zero-angular-moment-observers
(ZAMOs). Recently, this approach has been extended by Stuchlik et al.\cite{stuchlik_light_2018} to calculate shadow of Kerr-de Sitter black
holes. Alternatively, Grenzebach et al. \cite{grenzebach_photon_2014} firstly used Carter's frame \cite{bini_gyroscope_2017}
to calculate shadow of Kerr-de Sitter black holes for observers located at finite distance. This approach
has also been  used in other space-time geometries  \cite{eiroa_shadow_2018,haroon_shadow_2019}.
Thus, there are mainly two approaches for calculating the shadow of
rotating black holes in non-asymptotic flat space-time. The difference is
choice of orthonormal tetrads. Unfortunately, the two approaches lead to different shapes of shadow for the
observers in near regions. 

In this paper, we present a new approach to calculate shadow of rotating
black holes for observers located at finite distance without introducing 
tetrads. The powerful tools we used are astrometric observables, namely,
observed angle between two incident light rays in the celestial sphere. We
present analytical formulas for shadows of general rotating
black holes. Size and shape of Kerr-de Sitter black hole shadow are expressed as
functions of distance. For distant observers, our results are
consistent with previous works \cite{bardeen_timelike_1973,grenzebach_photon_2014}. For observers near the rotating black hole, our results are closed to Grenzebach et al.'s one \cite{grenzebach_photon_2014}.

This paper is organized as follows. In section \ref{II}, we utilize formula of
astrometric observable to calculate the shadow of spherical black holes. Familiar results of Synge \cite{synge_escape_1966} is recovered. In section
\ref{III}, we calculate the shadow of a general rotating
black hole in terms of astrometric observables. For given light rays from photon region, we present analytic
formula of size and shape of shadows. In
section \ref{IV}, we apply the new approach to Kerr-de Sitter black holes and
study how the size and shape of shadow changes with the location of observers. In
section \ref{V}, we compare our results with Bardeen's \cite{bardeen_timelike_1973} and Grenzebach et al.'s \cite{grenzebach_photon_2014} approaches.
Finally, conclusions and discussions are summarised in section \ref{VI}.

\section{Astrometric observables and shadow of Spherical black
  holes}\label{II}

It is well known that the directed observable in astrometry is angle between two
incident light rays $w$ and $k$ \cite{Soffel:2019aoq,lebedev_influence_2013},
\begin{equation}
	\cos \psi \equiv \frac{\gamma^* w \cdot \gamma^* k}{|
		\gamma^* w |   | \gamma^* k |} ~, \label{3}
\end{equation}
where the inner product is defined by space-time metric $g_{\mu \nu}$ and
$\gamma^*$ is projector for given 4-velocity $u$, namely
$\gamma^{\mu}_{\nu} = \delta^{\mu}_{\nu} + u^{\mu} u_{\nu}$. We can rewrite
Eq.~(\ref{3}) as
\begin{equation}
	\cos \psi  =  \frac{g_{\mu \nu} (\gamma^{\mu}_{\sigma} w^{\sigma})		(\gamma^{\nu}_{\rho} k^{\rho})}{\sqrt{\gamma_{\alpha \beta} w^{\alpha} w^{\beta}} \sqrt{\gamma_{\lambda \kappa} k^{\kappa} k^{\lambda}}} =  \frac{w \cdot k}{(u \cdot w) (u \cdot k)} + 1 ~ , \label{4}
\end{equation}
or
\begin{eqnarray}
	\cot \psi & = & {\rm sign} \left( \frac{\pi}{2} - \psi \right) \sqrt{- 1 -
		\left( \frac{1}{w \cdot k} \right) \frac{(u \cdot w)^2 (u \cdot k)^2}{(w
			\cdot k) + 2 (u \cdot w) (u \cdot k)}} ~ . \label{5}
\end{eqnarray}

\begin{figure}[ht]
	\centering
	{\includegraphics[width=0.7\linewidth]{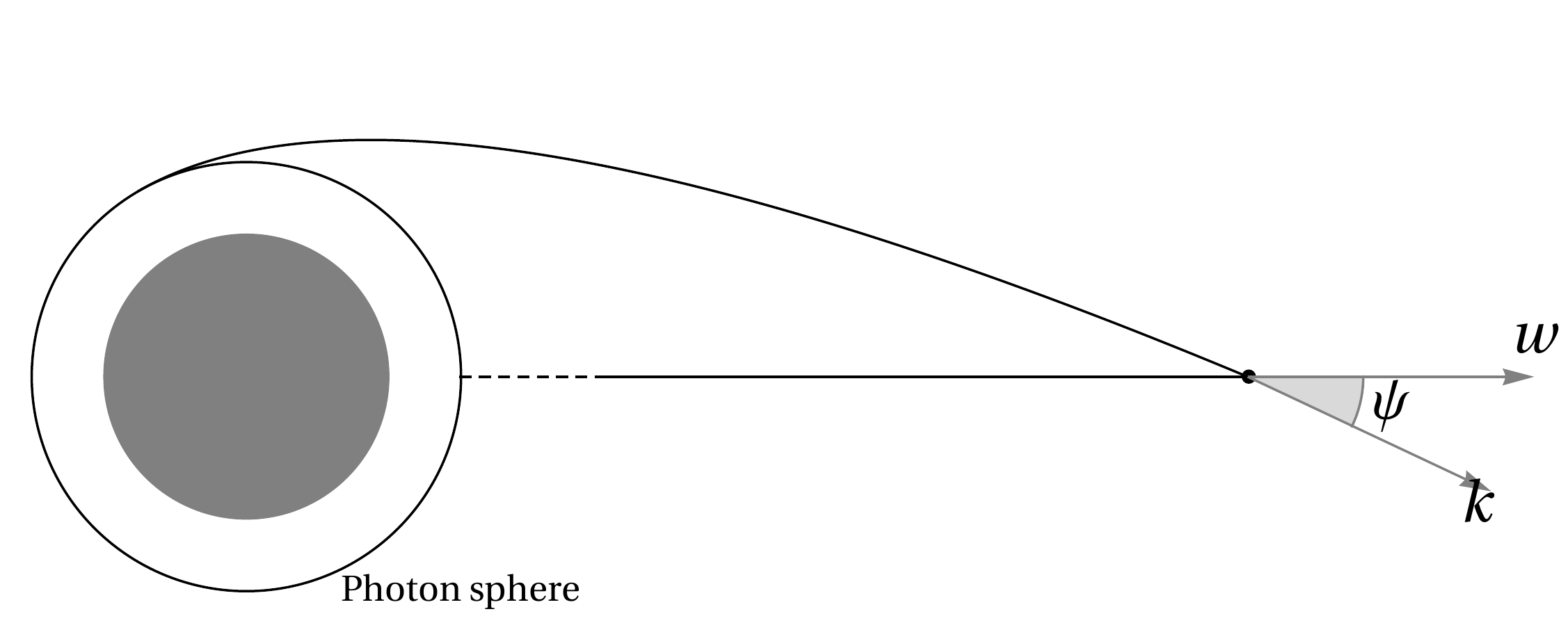}}
	\caption{Schematic diagram of measurement of spherical black hole shadow in astrometry.  The $\psi$ is angular radius of shadow. The $k$ is light ray from photon sphere, and $w$ is an auxiliary null vector. \label{Fig1}}
\end{figure}

For spherical black holes, the metric can be expressed as
\begin{equation}
	{\rm d} s^2 = g_{00} ({\rm d} x^0)^2 + g_{11} ({\rm d} x^1)^2 + g_{22} ({\rm d}
	x^2)^2 + g_{33} ({\rm d} x^3)^2 ~ .
\end{equation}
Here, we would show a general way to calculate size of spherical
black hole shadow for static observables, $u = \frac{1}{\sqrt{- g_{00}}}
	\partial_0$. In the framework of astrometry, schematic diagram is shown in
Figure~\ref{Fig1}. The angular radius of shadow $\psi$ is determined by a straight light ray
$w$ from centre of black holes and a bending light ray $k$ from photon sphere.
For the radial null geodesic $w = (w^0, w^1, 0, 0)$ and the observed light ray
$k = (k^0, k^1, 0, k^3)$, Eq.~(\ref{4}) becomes as
\begin{eqnarray}
	\cos \psi & = & \frac{g_{00} k^0 w^0 + g_{11} k^1 w^1}{- g_{00}   k^0
		w^0} + 1 \nonumber\\
	& = & \sqrt{- \frac{g_{11}}{g_{00}}} \frac{k^1}{k^0} ~ . \label{7}
\end{eqnarray}
It's consistent with the formula of angular radius of shadow proposed by Cunningham
\cite{cunningham_optical_1975} and used in recent Ref.~\cite{stuchlik_optical_2004}. Additionally, we can
rewrite Eq.~(\ref{7}) as
\begin{eqnarray}
	\cot \psi & = & {\rm sign} \left( \frac{\pi}{2} - \alpha \right) \sqrt{- 1
		- \left( \frac{1}{g_{00} k^0 w^0 + g_{11} k^1 w^1} \right) \frac{(g_{00}
			k^0 w^0)^2}{g_{00} k^0 w^0 + g_{11} k^1 w^1 - 2 g_{00}   k^0
			w^0}} \nonumber\\
	& = & \sqrt{\frac{g_{11}}{g_{33}}} \frac{k^1}{k^3} ~ . \label{8}
\end{eqnarray}
It's consistent with the formula proposed by Synge \cite{synge_escape_1966} and used in
recent Refs.~\cite{perlick_black_2018,tsupko_first_2019}. It shows that  
Eqs.~(\ref{7}) and (\ref{8}) in previous works  \cite{synge_escape_1966,cunningham_optical_1975} can derive from formula of
astrometric observables (Eq.~(\ref{3})). In the derivation, the straight light
ray $w$ functions as an auxiliary null vector. The final results have no
relevance with $w$. The formula of angular radius also can be checked by calculating angular diameter of shadow, which is exactly twice of the angular
radius $\psi$.

On the other side, 
 Eq.~(\ref{8}) can
be expressed in terms of orthonormal tetrads $k^{(a)} =
	e^{(a)}_{\hspace{0.5em} \mu} k^{\mu}$, where $e^{(a)}_{\hspace{0.5em} \mu} =
	{\rm diag} \left( \sqrt{- g_{00}}, \sqrt{g_{11}}, \sqrt{g_{22}},
	\sqrt{g_{33}} \right)$. From the geometrical intuitive, angular radius should
take form of
\begin{equation}
	\psi = \cot^{- 1} \left( \frac{k^{(1)}}{k^{(3)}} \right) ~ .
\end{equation}
In this sense, the angular radius is  angle between light ray $k$ and
radial coordinate line.

\section{Astrometric observable for shadow of rotating black holes}\label{III}

For rotating black holes, we can consider a general metric in the form,
\begin{equation}
	{\rm d} s^2 = - g_{00} {\rm d} t^2 + g_{11} {\rm d} r^2 + g_{22} {\rm d}
	\theta^2 + g_{33} {\rm d} \phi^2 + 2 g_{03} {\rm d} t {\rm d} \phi ~. \label{10}
\end{equation}
One might find troubles in understanding angular radius of shadow in
terms of orthonormal tetrads with geometrical intuitive. The problem is that
non-diagonal component of metric would lead to more than one reasonable
choices of orthonormal tetrads, but none of them seems preferred than others.

In the example of Kerr black holes, there are three orthonormal tetrad families,
namely, frame of ZAMOs, Carter's frame and static frame \cite{bini_gyroscope_2017}. In
static frame, the 0-component of tetrad $e_0$ is along the direction of
coordinate time $\partial_t$. The non-diagonal component of metric would lead to that 4-velocity $p_{(\phi)}$ (in tetrad) and $p_{\phi}$ (in coordinate
basis $\partial_{\mu}$) are not in the same direction. Thus, the static frame seems not suited for
understanding the angular radius with geometrical intuitive as that for
spherical black holes. Maybe, due to this consideration, Bardeen firstly
considered the shadow of Kerr black holes with respect to ZAMOs. In this local
frame, the 0-component of tetrad $e_0$ is not adapted to a static observer,
while spatial components of 4-velocity $p_{(i)}$ are in the same direction of $p_i$. It
 gives a well-understood formulation for angular radius of rotating black hole shadow
in the framework of tetrads,
\begin{eqnarray}
	\alpha_{\rm {Bard}} & = & - \left. \frac{p_{(\phi)}}{p_{(t)}}
	\right|_{\rm {ZAMO}} ~ , \\
	\beta_{\rm {Bard}} & = &  \left. \frac{p_{(\theta)}}{p_{(t)}}
	\right|_{\rm {ZAMO}} ~ .
\end{eqnarray}
Carter's frame is fundamentally important to properties of geodesic
equations and curvature tensors \cite{bini_gyroscope_2017}. Grenzebach et al. \cite{grenzebach_photon_2014} firstly
used this orthonormal frame to calculate shadow of Kerr-like black holes. However, in
near zone from black holes, one might find shapes of Kerr black holes are
different with different choice of orthonormal tetrads \cite{bardeen_timelike_1973,grenzebach_photon_2014}. 

In this section, we would alternatively use formula of
astrometric observables to calculate shadow of rotating black hole
(Eq.~(\ref{10})). For the sake of intuitive, we consider inclination angle $\theta = 0$
and $\theta = \frac{\pi}{2}$ as representative cases. 
For rotating black holes, there is not a straight null
geodesic as auxiliary vector in general. In the case of $\theta = 0$, we can
use the same calculation as that for spherical black hole
shadow. While, in the case of $\theta = \frac{\pi}{2}$, we should use geometric
trigonometry in celestial sphere.

\subsection{Inclination angle $\theta = 0$}

\begin{figure}[ht]
	\centering
	\includegraphics[width=0.6\linewidth]{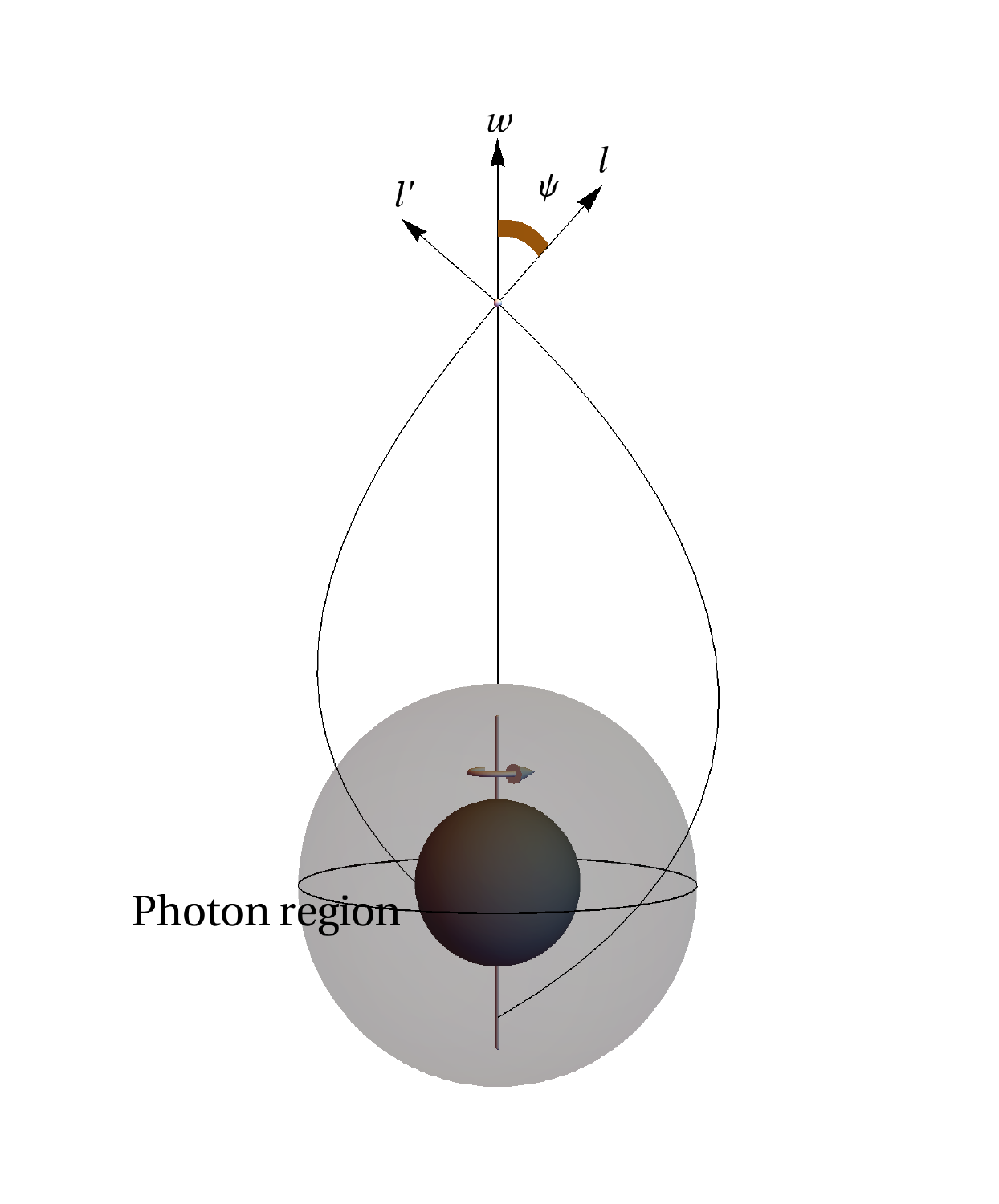}
	\caption{Schematic diagram of measurement of rotating black hole shadow in astrometry. The observers are located at inclination angle $\theta=0$. The $\psi$ is angular radius of shadow. The $l$ is light ray from photon region, and $w$ is an auxiliary null vector.}
	\label{Fig2}
\end{figure}

As schematic diagram shown in Figure~\ref{Fig2}, we consider a straight null curve $w = (w^0, w^0, 0, 0)$ along rotation axis of the rotating black hole and a
bending light ray $l = (l^0, l^1, l^2, l^3)$ from photon region. Without spherical symmetric, we can not simply set the third component of light ray $l^2 =
	0$. In this case, the angular radius $\psi$ can be given by Eq.~(\ref{5}),
\begin{eqnarray}
	\cot \psi & = & {\rm sign} \left( \frac{\pi}{2} - \psi \right) \sqrt{- 1 -
		\left( \frac{1}{w \cdot l} \right) \frac{(u \cdot w)^2 (u \cdot l)^2}{(w
			\cdot l) + 2 (u \cdot w) (u \cdot l)}} \nonumber\\
	& = & {\rm sign} \left( \frac{\pi}{2} - \psi \right)
	\sqrt{\frac{g_{11}}{g_{22} \left( \frac{l^2}{l^1} \right)^2 + \left( g_{33}
			- \frac{g_{03}^2}{g_{00}} \right) \left( \frac{l^3}{l^1} \right)^2}} ~ . \label{13}
\end{eqnarray}
For $g_{03} = 0$, it reduces to the formula for spherical black holes. One might find that  Eq.~(\ref{13}) can not simply read in terms of the tetrads mentioned above \cite{bini_gyroscope_2017}.

\subsection{Inclination angle $\theta = \frac{\pi}{2}$}
\begin{figure}[ht]
	\centering
	\includegraphics[width=0.8\linewidth]{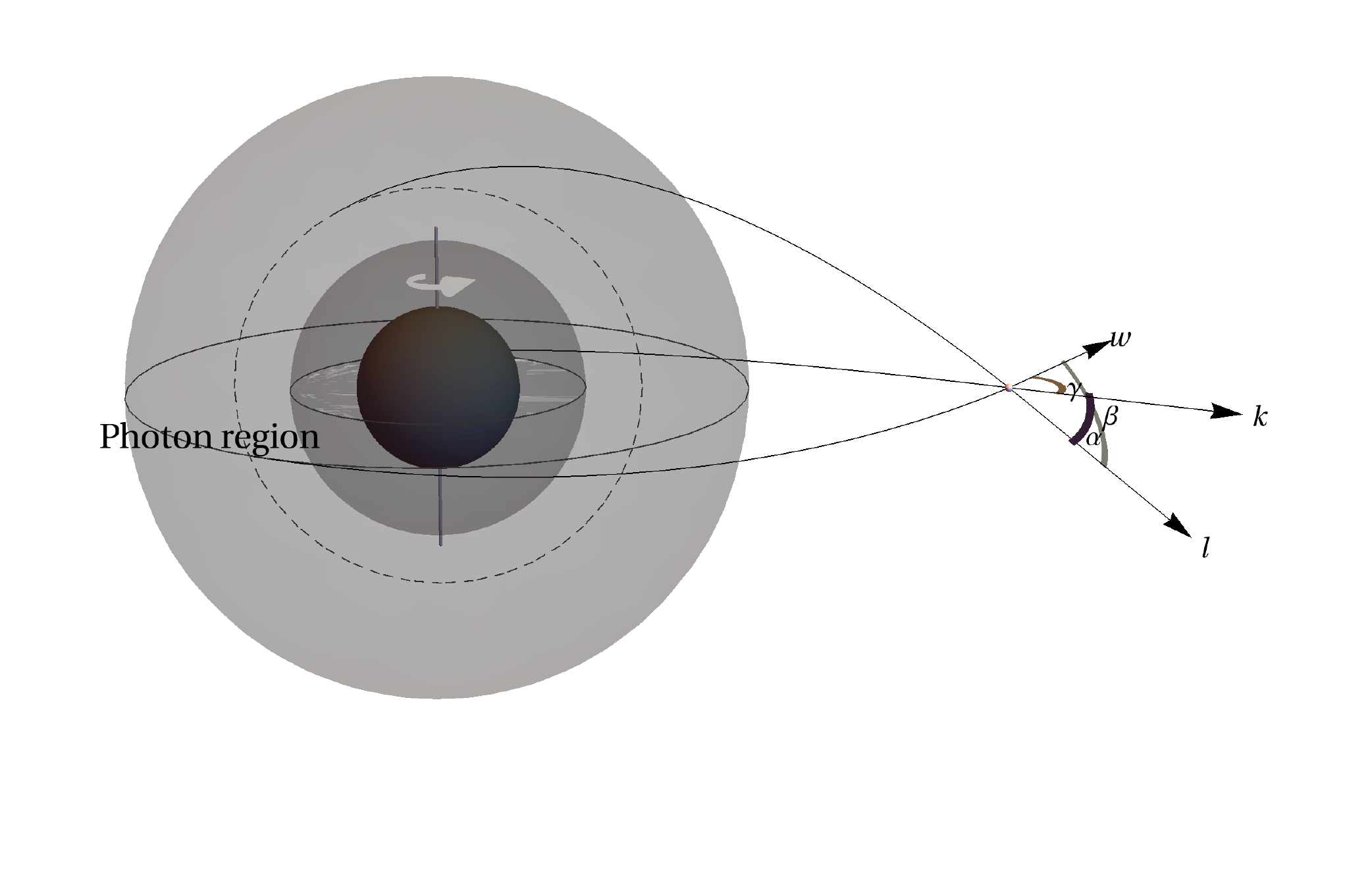}
	\caption{Schematic diagram of measurement of rotating black hole shadow in astrometry. The observers are located at inclination angle $\theta=\frac{\pi}{2}$.  The $k$, $w$ and $l$ are light rays from photon region. The $\gamma$, $\alpha$, $\beta$ are angles between $k$ and $w$, $l$ and $k$, $l$ and $w$, respectively.}
	\label{Fig3}
\end{figure}

In the case of observers located at equatorial plane of a black hole, the
calculation of rotating black hole shadows turns to be more tricky. As schematic diagram shown in Figure~\ref{Fig3}.
we consider two bending light rays $k$ and $w$ in the equatorial plane. Due to
frame dragging effect, the counter-rotating light ray $k$ has a lower angular
velocity than that of light ray $w$. In astrometry, we can measure the angular
distance $\gamma$ between $k$ and $w$ in the celestial sphere, which is formulated as
\begin{eqnarray}
	\cot \gamma & = & {\rm sgin} \left( \frac{\pi}{2} - \gamma \right) \sqrt{-
		1 - \left( \frac{1}{k \cdot w} \right) \frac{(u \cdot k)^2 (u \cdot w)^2}{(k
			\cdot w) + 2 (u \cdot k) (u \cdot w)}} \nonumber\\
	& = & {\rm sign} (k, w) \sqrt{\frac{\left( g_{11} \frac{k^1 w^1}{k^3 w^1 -
				w^3 k^1} + \left( g_{33} - \frac{g_{03}^2}{g_{00}} \right) \frac{k^3
				w^3}{k^3 w^1 - w^3 k^1} \right)^2}{g_{11} \left( g_{33} -
			\frac{g_{03}^2}{g_{00}} \right)}} \nonumber\\
	& \equiv & {\rm sign} (k, w) \sqrt{\frac{\left(
			\frac{g_{11}}{\mathcal{K}-\mathcal{W}} + \left( g_{33} -
			\frac{g_{03}^2}{g_{00}} \right) \frac{1}{\frac{1}{\mathcal{W}} -
				\frac{1}{\mathcal{K}}} \right)^2}{g_{11} \left( g_{33} -
			\frac{g_{03}^2}{g_{00}} \right)}} ~,  \label{14}
\end{eqnarray}
where $\mathcal{K} \equiv \frac{k^3}{k^1}, \mathcal{W} \equiv
	\frac{w^3}{w^1}$, and
\begin{equation}
	{\rm sign} (k, w) = {\rm sign} \left( g_{11} + \left( g_{33} -
	\frac{g_{03}^2}{g_{00}} \right) \mathcal{K}\mathcal{W} \right) ~ .
\end{equation}
Because of axisymmetry of rotating black holes, here, we can set $k = (k^0,
	k^1, 0, k^3)$, $w = (w^0, w^1, 0, w^3)$ for simplicity.

As we known, location in the celestial sphere can be determined by two
parameters. For an light ray $l$ from photon region, we determine location of $l$ by parameters $\alpha$ and
$\beta$. The $\alpha$ is  angle between $l$ and $k$, and the $\beta$ is 
angle between $l$ and $w$. The angular distances $\alpha$ and $\beta$ can be expressed as
\begin{eqnarray}
	\cot \alpha & = & {\rm sign} \left( \frac{\pi}{2} - \alpha \right) \sqrt{-
		1 - \left( \frac{1}{k \cdot l} \right) \frac{(u \cdot k)^2 (u \cdot l)^2}{(k
			\cdot l) + 2 (u \cdot k) (u \cdot l)}} \nonumber\\
	& = & {\rm sign} (k, l) \sqrt{\frac{\left( g_{11}
			\frac{1}{\mathcal{K}-\mathcal{L}_3} + \left( g_{33} -
			\frac{g_{03}^2}{g_{00}} \right) \frac{1}{\frac{1}{\mathcal{L}_3} -
				\frac{1}{\mathcal{K}}} \right)^2}{g_{22} \left( g_{11} \left(
			\frac{\mathcal{L}_2}{\mathcal{K}-\mathcal{L}_3} \right)^2 + \left( g_{33} -
			\frac{g_{03}^2}{g_{00}} \right) \left( \frac{\mathcal{L}_2}{1 -
				\frac{\mathcal{L}_3}{\mathcal{K}}} \right)^2 \right) + g_{11} \left( g_{33}
			- \frac{g_{03}^2}{g_{00}} \right)}} ~, \label{16}
\end{eqnarray}
and
\begin{eqnarray}
	\cot \beta & = & {\rm sign} \left( \frac{\pi}{2} - \beta \right) \sqrt{- 1
		- \left( \frac{1}{w \cdot l} \right) \frac{(u \cdot w)^2 (u \cdot l)^2}{(w
			\cdot l) + 2 (u \cdot w) (u \cdot l)}} \nonumber\\
	& = & {\rm sign} (w, l) \sqrt{\frac{\left( g_{11}
			\frac{1}{\mathcal{W}-\mathcal{L}_3} + \left( g_{33} -
			\frac{g_{03}^2}{g_{00}} \right) \frac{1}{\frac{1}{\mathcal{L}_3} -
				\frac{1}{\mathcal{K}}} \right)^2}{g_{22} \left( g_{11} \left(
			\frac{\mathcal{L}_2}{\mathcal{W}-\mathcal{L}_3} \right)^2 + \left( g_{33} -
			\frac{g_{03}^2}{g_{00}} \right) \left( \frac{\mathcal{L}_2}{1 -
				\frac{\mathcal{L}_3}{\mathcal{W}}} \right)^2 \right) + g_{11} \left( g_{33}
			- \frac{g_{03}^2}{g_{00}} \right)}} ~, \label{17}
\end{eqnarray}
where $\mathcal{L}_2 \equiv \frac{l^3}{l^1}, \mathcal{L}_3 \equiv
	\frac{l^3}{l^1}$.

\begin{figure}[ht]
	\centering
	\includegraphics[width=0.7\linewidth]{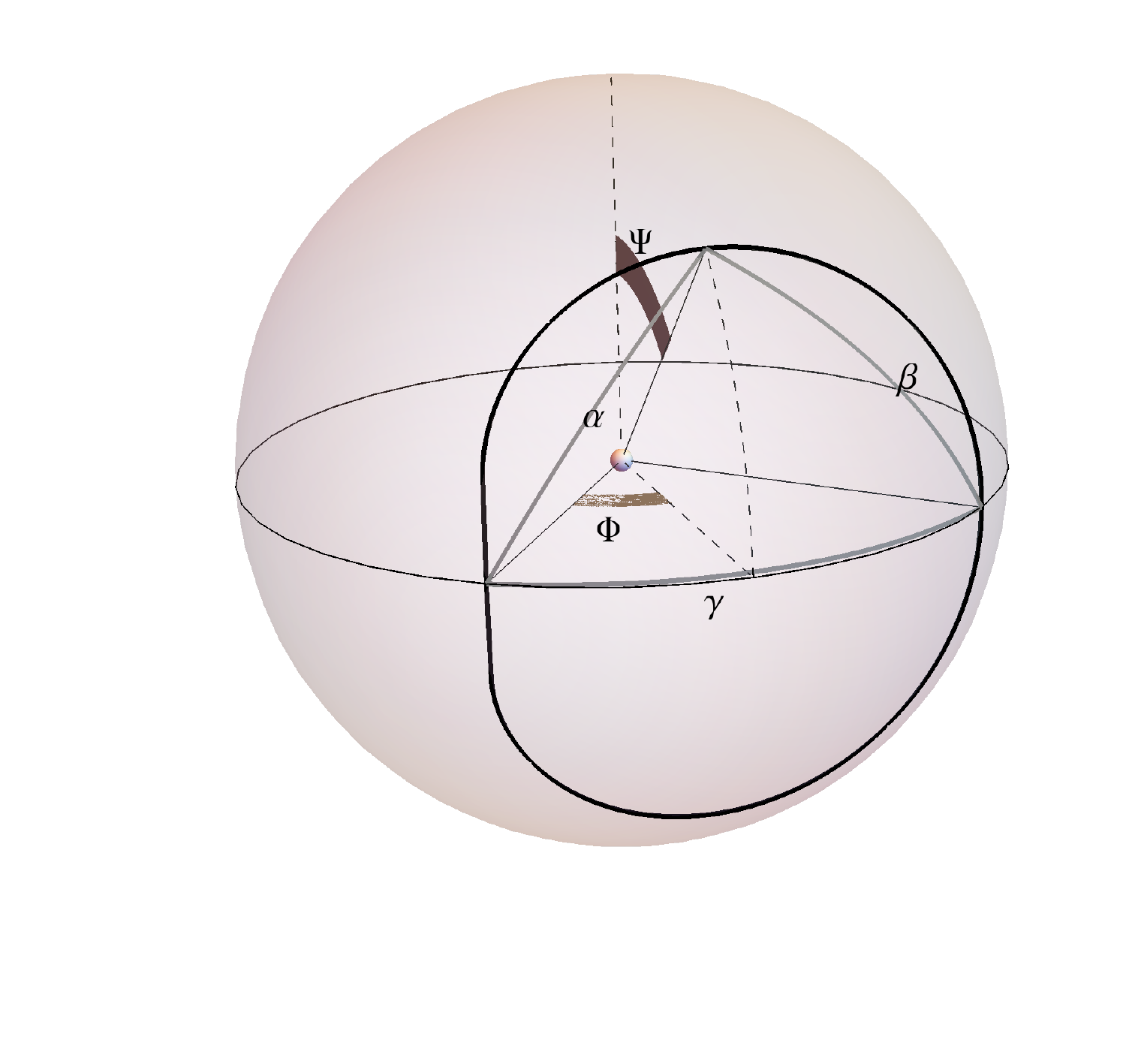}
	\caption{Schematic diagram for the shadow of rotating black holes in view of observers. The $\alpha$, $\beta$ and $\gamma$ are the angles shown in Figure~\ref{Fig3}. The $\Psi$ and $\Phi$ are the celestial coordinates in the view of observers. The distorted circle in the celestial sphere is the boundary of shadow of a rotating black hole. Here, we use the shadow of Kerr black holes for instance. }
	\label{Fig4}
\end{figure}

In Figure~\ref{Fig4}, we present schematic
diagram for  shadow of rotating black holes in view of observers. The boundary of
shadow is determined by all the light rays $l$ received by observers.
And celestial coordinates $(\Psi, \Phi)$ of shadow can be expressed in terms of $\alpha$, $\beta$ and $\gamma$,
\begin{eqnarray}
	{\cos A} & = & \frac{\cos \alpha - \cos \beta \cos \gamma}{\sin \beta
		\sin \gamma} ~,\\
	\frac{\sin \beta}{\sin \frac{\pi}{2}} & = & \frac{\sin \left(\frac{\pi}{2}-\Psi\right)}{\sin A}~,
	\\
	\cos \beta & = & \cos \left( \frac{\pi}{2} - \Psi \right) \cos (\gamma -
	\Phi) + \sin \left( \frac{\pi}{2} - \Psi \right) \cos (\gamma - \Phi) \cos
	\left( \frac{\pi}{2} \right) ~.
\end{eqnarray}
Then, we have
\begin{eqnarray}
	\Psi & = & \frac{\pi}{2} - \arcsin \left( \sin \beta \sqrt{1 - \left(
		\frac{\cos \alpha - \cos \beta \cos \gamma}{\sin \beta \sin \gamma}
		\right)^2} \right) ~, \label{21}\\
	\Phi & = & \gamma - \arccos \left( \frac{\cos \beta}{\cos \left(
		\frac{\pi}{2} - \Psi \right)} \right) ~. \label{22}
\end{eqnarray}

\begin{figure}[ht]
	\centering
	\includegraphics[width=0.7\linewidth]{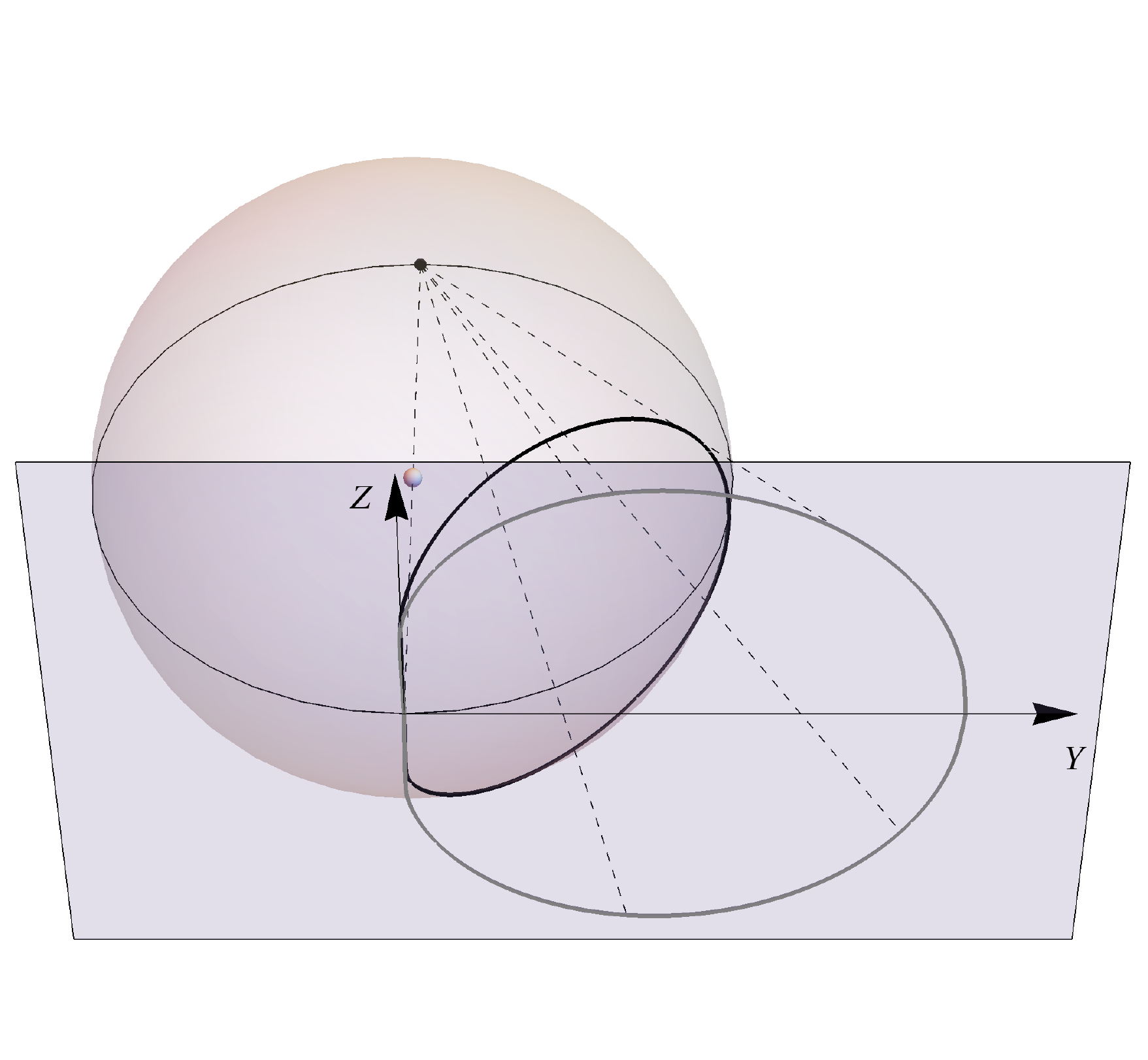}
	\caption{Schematic diagram for stereographic projection. Here, we use Kerr black holes for instance.}
	\label{Fig5}
\end{figure}

The most interesting part of rotating black hole shadow is lying on its shape.
Because of frame dragging effect on propagating light rays, the shape of
shadow is usually distorted from a circle in the view of observers. In order
 to describe clearly the shape of shadow in 2D-plane, we use a stereographic projection for the celestial
coordinates,
\begin{eqnarray}
	Y_{\rm {sh}} & = & \frac{2 \sin \Phi \sin \Psi}{1 + \cos \Phi \sin \Psi}
	~, \label{23} \\
	Z_{\rm {sh}} & = & \frac{2 \cos \Psi}{1 + \cos \Phi \sin \Psi} ~ . \label{24}
\end{eqnarray}
The schematic diagram for the stereographic projection is presented in Figure~\ref{Fig5}.
As a circle in celestial sphere is mostly projected to a circle on
2D-plane, the shape of shadow beyond a circle in 2D-plane can involve
property of space-time beyond spherical black holes.

In this section, we calculate shadow of a general rotating black hole by making use of the formula of
astrometric observables. 
In Eqs.~(\ref{13}), (\ref{14}), (\ref{16}) and (\ref{17}) , the light rays $k$, $w  $ and $l$ are moving along null geodesics. And the
integral constants of these null vectors are determined by the photon region of rotating black holes.
 Thus, for the sake of intuitive, we would apply our formula to Kerr-de Sitter black holes in next section.

\section{Application in Kerr-de Sitter Black holes}\label{IV}

The space-time of Kerr-de Sitter black holes is non-asymptotically flat. The
calculation of shadow for an observer located at spatial infinity wouldn't be valid any
more. By introducing orthonormal tetrads, Grenzebach et al. \cite{grenzebach_photon_2014}  discussed the shadow for
observers fixed in finite distance. In this paper, we also aim
at this situation. Without introducing tetrads, we use formula of astrometric observables instead,

The metric of Kerr-de Sitter black holes \cite{grenzebach_photon_2014,haroon_shadow_2019,konoplya_quasinormal_2011} is given by
\begin{equation}
	{\rm d} s^2 = - \frac{\Delta_r}{I^2 \Sigma} ({\rm d} t -  {a \sin}^2 \theta
	{\rm d} \phi)^2 + \frac{\Delta_{\theta} \sin^2 \theta}{I^2 \Sigma} (a {\rm d}
	t - (r^2 + a^2) {\rm d} \phi)^2 + \frac{\Sigma}{\Delta_r} {\rm d} r^2 +
	\frac{\Sigma}{\Delta_{\theta}} {\rm d} \theta^2~, \label{25}
\end{equation}
where
\begin{eqnarray}
	\Delta_r (r) & = & - \frac{1}{3} \Lambda r^2 (r^2 + a^2) + r^2 - 2 M
	r + a^2 ~, \\
	\Delta_{\theta} (\theta) & = & 1 + \frac{1}{3} \Lambda a^2 \cos^2 \theta ~,\\
	I & = & 1 + \frac{1}{3} \Lambda a^2 ~,\\
	\Sigma (r, \theta) & = & r^2 + a^2 \cos^2 \theta ~.
\end{eqnarray}
The $M$ is black hole mass, $a$ is spin parameter and $\Lambda$ is cosmological constant. From the metric (Eq.~(\ref{25})), one can obtain 4-velocities of light rays via
Hamiltonian-Jacobi method (see, for example, \cite{bardeen_timelike_1973,grenzebach_photon_2014,haroon_shadow_2019}),
\begin{eqnarray}
	\Sigma p^t & = & I^2 E \left( \frac{(r^2 + a^2 - a \lambda) (r^2 +
		a^2)}{\Delta_r} + \frac{a (\lambda - {a \sin}^2 \theta)}{\Delta_{\theta}}
	\right) ~,\\
	(\Sigma p^r)^2 & = & R (r) ~,\\
	(\Sigma p^{\theta})^2 & = & \Theta (\theta) ~,\\
	\Sigma p^{\phi} & = & I^2 E \left( \frac{a (r^2 + a^2) - a^2
		\lambda}{\Delta_r} + \frac{\lambda - {a \sin}^2 \theta}{\Delta_{\theta}
		\sin^2 \theta} \right) ~,
\end{eqnarray}
where
\begin{eqnarray}
	R (r) & = & E^2 (I^2 (r^2 + a^2 - a \lambda)^2 - \Delta_r \kappa) ~,\\
	\Theta (\theta) & = & E^2 \left( \Delta_{\theta} \kappa - \frac{I^2 (\lambda
	- {a \sin}^2 \theta)^2}{\sin^2 \theta} \right) ~,
\end{eqnarray}
and
\begin{eqnarray}
	\lambda & \equiv & \frac{L}{E}~, \\
	\kappa & \equiv & \frac{K}{E^2} ~.
\end{eqnarray}
The $L, E, K$ are integral constants from the null geodesic equations. The
photon region of Kerr-de Sitter space-time has been studied carefully in
Ref.~\cite{grenzebach_photon_2014}. It's determined by unstable circle orbits formulated as
\begin{eqnarray}
	R (r_c) & = & 0~, \\
	\left. \frac{{\rm d} R (r)}{{\rm d} r} \right|_{r = r_c} & = & 0 ~,
\end{eqnarray}
which lead to
\begin{eqnarray}
	\lambda (r_c) & = & \left. \frac{1}{a} \left( r^2 + a^2 - \frac{4 r
		\Delta_r}{\Delta'_r} \right) \right|_{r = r_c}~, \\
	\kappa (r_c) & = & \left. \frac{16 I^2 r^2 \Delta_r}{(\Delta_r')^2}
	\right|_{r = r_c} ~,
\end{eqnarray}
where $r_c$ is the location of photon region. The range of $r_c$ is determined
by $\Theta (\theta) \geqslant 0$, namely,
\begin{equation}
	((4 r \Delta_r - \Sigma \Delta'_r)^2 - 16 a^2 r^2 \Delta_r \Delta_{\theta}
	\sin^2 \theta)_{r = r_c} \leqslant 0~ \label{42}.
\end{equation}
Here, $r_{c -}$ and $r_{c +}$ are minimum and maximum radial position of
photon region outside of inner horizon. If limiting the null vectors from the
photon region, one can regard $p^{\mu}$ as function of $x^{\mu}$, $E$ and
$r_c$. One should note that $\theta$ is a coordinate of observers. As shown
in Figure~\ref{Fig2} and \ref{Fig3}, the photon region is different for different locations of
observers.

\subsection{Sizes of shadow}

For observers located at inclination angle $\theta = 0$, the Eq.~(\ref{42}) can be
rewritten as
\begin{equation}
	(4 r \Delta_r - (r^2 + a^2) \Delta'_r)_{r = r_c} = 0~.
\end{equation}
In this case, $r_c = r_{c -} = r_{c +} \equiv r_{c 0}$. Using Eq.~(\ref{13}), one can obtain angular radius of Kerr-de Sitter black hole shadow in the form,
\begin{equation}
	\cot \psi = \sqrt{\frac{I^2 (r^2 + a^2 - a \lambda_0)^2 - \Delta_r
			\kappa_0}{\lambda_0 a   I^2 (2 r^2 + 2 a^2 - a \lambda_0) + \Delta_r
			\kappa_0}}~,
\end{equation}
where $\lambda_0 \equiv \lambda (r_{c 0})$ and $\kappa_0 \equiv (r_{c 0})$.
Here, we only consider shadow in the view of observers located outside of
the photon region.

\begin{figure}[ht]
	\centering
	\includegraphics[width=1\linewidth]{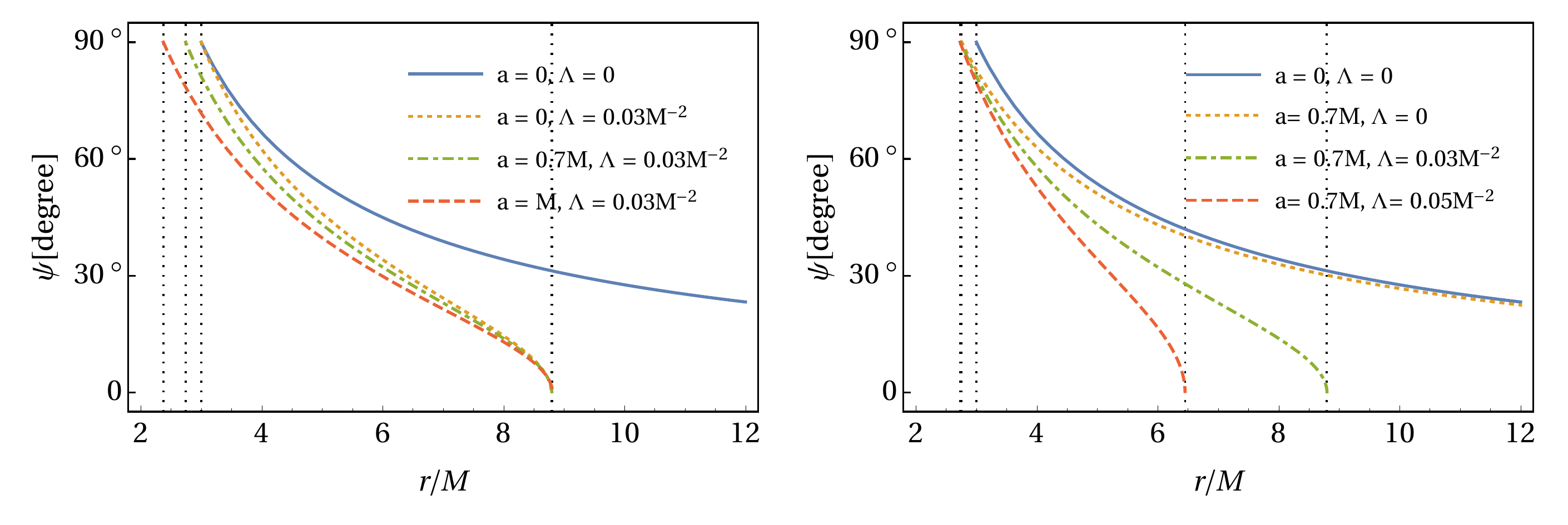}
	\caption{Angular radius as function of distance from rotating black holes for selected parameters. The observers are located at inclination angle $\theta=0$. The vertical dotted lines are outer boundaries and cosmological horizons. Left panel: Angular radius as function of distance for different spin parameters. Right panel: Angular radius as function of distance for different cosmological constants. }
	\label{Fig6}
\end{figure}

In Figure~\ref{Fig6}. we present the angular radius as function of distance from
central black hole. The $r_{c 0}$ is 
smaller than radius of photon sphere of Schwarzschild black holes.  In the left panel, it shows that
angular radius decreases with spin parameter $a$. And in the right panel, the
angular radius also decreases with cosmological constant $\Lambda$. Among these black holes, the size of
Schwarzschild black hole shadow is the largest.

For observers located at inclination angle $\theta = \frac{\pi}{2}$, we can
determine the range of $r_c$ from Eq.~(\ref{42}), which takes the form of
\begin{equation}
	((4 r \Delta_r - r^2 \Delta'_r)^2 - 16 a^2 r^2 \Delta_r)_{r = r_c} \leqslant  0~. \label{45}
\end{equation}
Namely, $r_{c -} \leqslant r_c \leqslant r_{c +}$. From Eq.~(\ref{14}), we get the angular
diameter $\gamma$ in the Kerr-de space-time,
\begin{equation}
	\cot \gamma = {\rm sign} \left( 1 + \frac{\Delta_r^2}{I^2 (\Delta_r - a^2)}
	\mathcal{K}\mathcal{W} \right) \left| \frac{I \sqrt{\Delta_r -
			a^2}}{\Delta_r} \frac{1}{\mathcal{K}-\mathcal{W}} + \frac{\Delta_r}{I
		\sqrt{\Delta_r - a^2}} \frac{1}{\frac{1}{\mathcal{W}} -
		\frac{1}{\mathcal{K}}} \right|~,
\end{equation}
where
\begin{eqnarray}
	\mathcal{K} & = & \left. \frac{p^{\phi}}{p^r} \right|_{r_c = r_{c -}} ~, \label{47}\\
	\mathcal{W} & = & \left. \frac{p^{\phi}}{p^r} \right|_{r_c = r_{c +}} ~, \label{48}
\end{eqnarray}
and
\begin{equation}
	\frac{p^{\phi}}{p^r} = \frac{I^2 \left( \frac{a (r^2 + a^2 - a
			\lambda)}{\Delta_r} + \lambda - a \right)}{\sqrt{I^2 (r^2 + a^2 - a
			\lambda)^2 - \Delta_r \kappa}}~. \label{49}
\end{equation}
Here,  $\frac{p^{\phi}}{p^r}$ is function of $r, r_c$.

\begin{figure}[ht]
	\centering
	\includegraphics[width=1\linewidth]{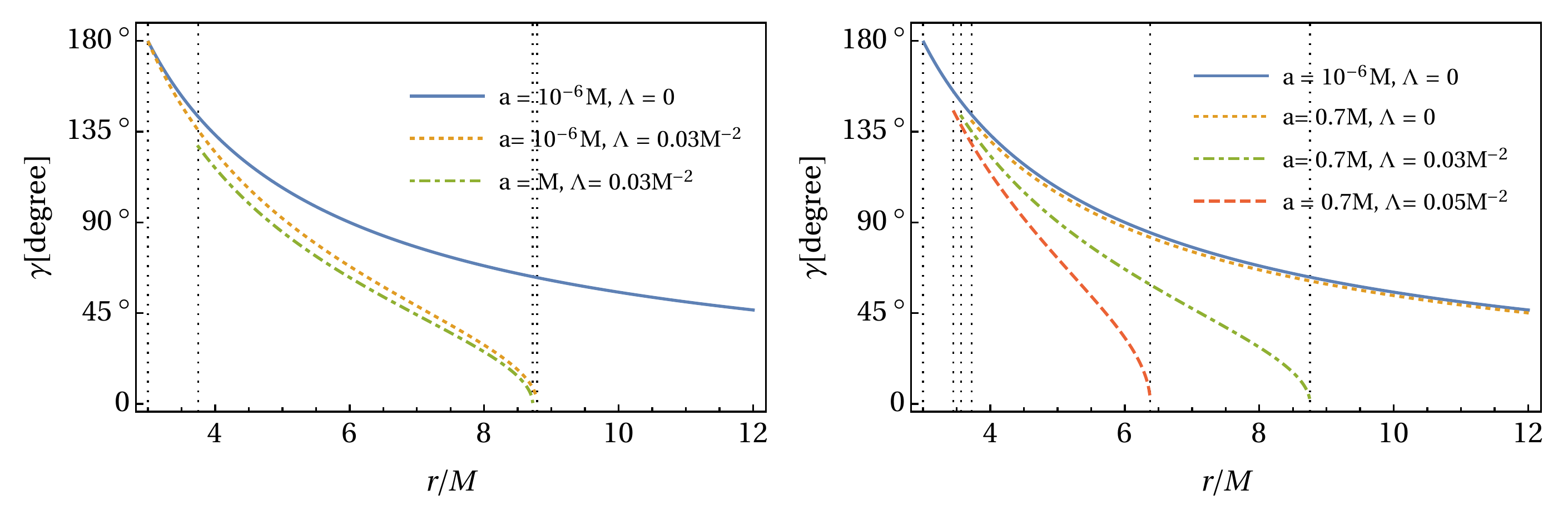}
	\caption{Angular diameter $\gamma$ as function of distance from rotating black holes for  selected parameters. The observers are located at inclination angle $\theta=\frac{\pi}{2}$. The vertical dotted   lines are outer boundaries and cosmological horizons. Left panel: Angular  diameter as function of distance for different spin parameters. Right panel: Angular  diameter as function of distance for different cosmological constants. }
	\label{Fig7}
\end{figure}

We plot angular diameter $\gamma$ as
function of distance $r$ in Figure~\ref{Fig7} . The outer boundary of photon region
$r_{c +}$ is larger than radius of photon sphere in Schwarzschild space-time.
In the left panel,  it shows that the spin parameters hardly affect the angular diameter. The right panel
shows that the angular diameter  decreases  with cosmological constant $\Lambda$.

We can conclude that the size of Kerr-de Sitter black hole shadow decreases
with spin parameter $a$ and cosmological constant $\Lambda$ and the shadow of
Schwarzschild black hole is the biggest among  Kerr-de Sitter black holes.

\subsection{Shapes of shadow}

In this part, we turn to shape of  Kerr-de Sitter black hole shadow as function of distance. 
The observers located at inclination angle $\theta = 0$ would see the shadow as
a perfect circle, while, the observers located at $\theta = \frac{\pi}{2}$
would find that the shadow is distorted. In this case of Kerr-de Sitter black holes, Eqs.~(\ref{16}) and (\ref{17}) read
\begin{eqnarray}
	\cot \alpha & = & {\rm sign} \left( 1 + \frac{\Delta_r^2}{I^2 (\Delta_r -
			a^2)} \mathcal{K}\mathcal{L}_3 \right) \frac{\left| \frac{I \sqrt{\Delta_r -
				a^2}}{\Delta_r} \frac{1}{\mathcal{K}-\mathcal{L}_3} + \frac{\Delta_r}{I
			\sqrt{\Delta_r - a^2}} \frac{1}{\frac{1}{\mathcal{L}_3} -
			\frac{1}{\mathcal{K}}} \right|}{\sqrt{1 + \frac{I^2 (\Delta_r -
				a^2)}{\Delta_r} \left( \frac{\mathcal{L}_2}{\mathcal{K}-\mathcal{L}_3}
			\right)^2 + \Delta_r \left( \frac{\mathcal{L}_2}{1 -
				\frac{\mathcal{L}_3}{\mathcal{K}}} \right)^2}}~,\\
	\cot \beta & = & {\rm sign} \left( 1 + \frac{\Delta_r^2}{I^2 (\Delta_r -
			a^2)} \mathcal{W}\mathcal{L}_3 \right) \frac{\left| \frac{I \sqrt{\Delta_r -
				a^2}}{\Delta_r} \frac{1}{\mathcal{W}-\mathcal{L}_3} + \frac{\Delta_r}{I
			\sqrt{\Delta_r - a^2}} \frac{1}{\frac{1}{\mathcal{L}_3} -
			\frac{1}{\mathcal{W}}} \right|}{\sqrt{1 + \frac{I^2 (\Delta_r -
				a^2)}{\Delta_r} \left( \frac{\mathcal{L}_2}{\mathcal{W}-\mathcal{L}_3}
			\right)^2 + \Delta_r \left( \frac{\mathcal{L}_2}{1 -
				\frac{\mathcal{L}_3}{\mathcal{W}}} \right)^2}}~,
\end{eqnarray}
where $\mathcal{W}, \mathcal{K}$ are given by Eqs.~(\ref{47}), (\ref{48}) and
\begin{eqnarray}
	\mathcal{L}_2 & \equiv & \left. \frac{p^{\theta}}{p^r} \right|_{r_c}~,
\end{eqnarray}
\begin{eqnarray}
	\mathcal{L}_3 & \equiv & \left. \frac{p^{\phi}}{p^r} \right|_{r_c}~,
\end{eqnarray}
and
\begin{equation}
	\frac{p^{\theta}}{p^r} = \pm \sqrt{\frac{\kappa - I^2 (\lambda - a)^2}{I^2
			(r^2 + a^2 - a \lambda)^2 - \Delta_r \kappa}} ~.
\end{equation}
The $r_{c -} \leqslant r_c \leqslant r_{c +}$, is determined by Eq.~(\ref{45}). In terms of the
angular distance $\alpha, \beta$ and $\gamma$, we can give the shape of
shadow on the projective plane $(X, Y)$. Namely, from Eqs. (\ref{21})--(\ref{24}), the
boundary of shadow is described by
\begin{eqnarray}
	Y_{\rm {sh}} & = & \frac{2 \cos \beta \sin \gamma - 2 \cot \gamma
		\sqrt{\sin^2 \gamma \sin^2 \beta + (\cos (\beta + \gamma) - \cos \alpha)
			(\cos (\beta - \gamma) - \cos \alpha)}}{1 + \cos \beta \cos \gamma +
		\sqrt{\sin^2 \gamma \sin^2 \beta + (\cos (\beta + \gamma) - \cos \alpha)
			(\cos (\beta - \gamma) - \cos \alpha)}}~, \label{55}\\
	Z_{\rm {sh}} & = & \frac{2 \csc\gamma\sqrt{(\cos \alpha - \cos (\beta + \gamma))
			(\cos (\beta - \gamma) - \cos \alpha)}}{1 + \cos \beta \cos \gamma +
		\sqrt{\sin^2 \gamma \sin^2 \beta + (\cos (\beta + \gamma) - \cos \alpha)
			(\cos (\beta - \gamma) - \cos \alpha)} } ~. \label{56}
\end{eqnarray}
On the projective plane $(Y, Z)$, the boundary of shadow is described by a
parametrized curve in terms of parameter $r_c$, which has been shown in Figure~\ref{Fig5}.

\begin{figure}[ht]
	\centering
	\includegraphics[width=1\linewidth]{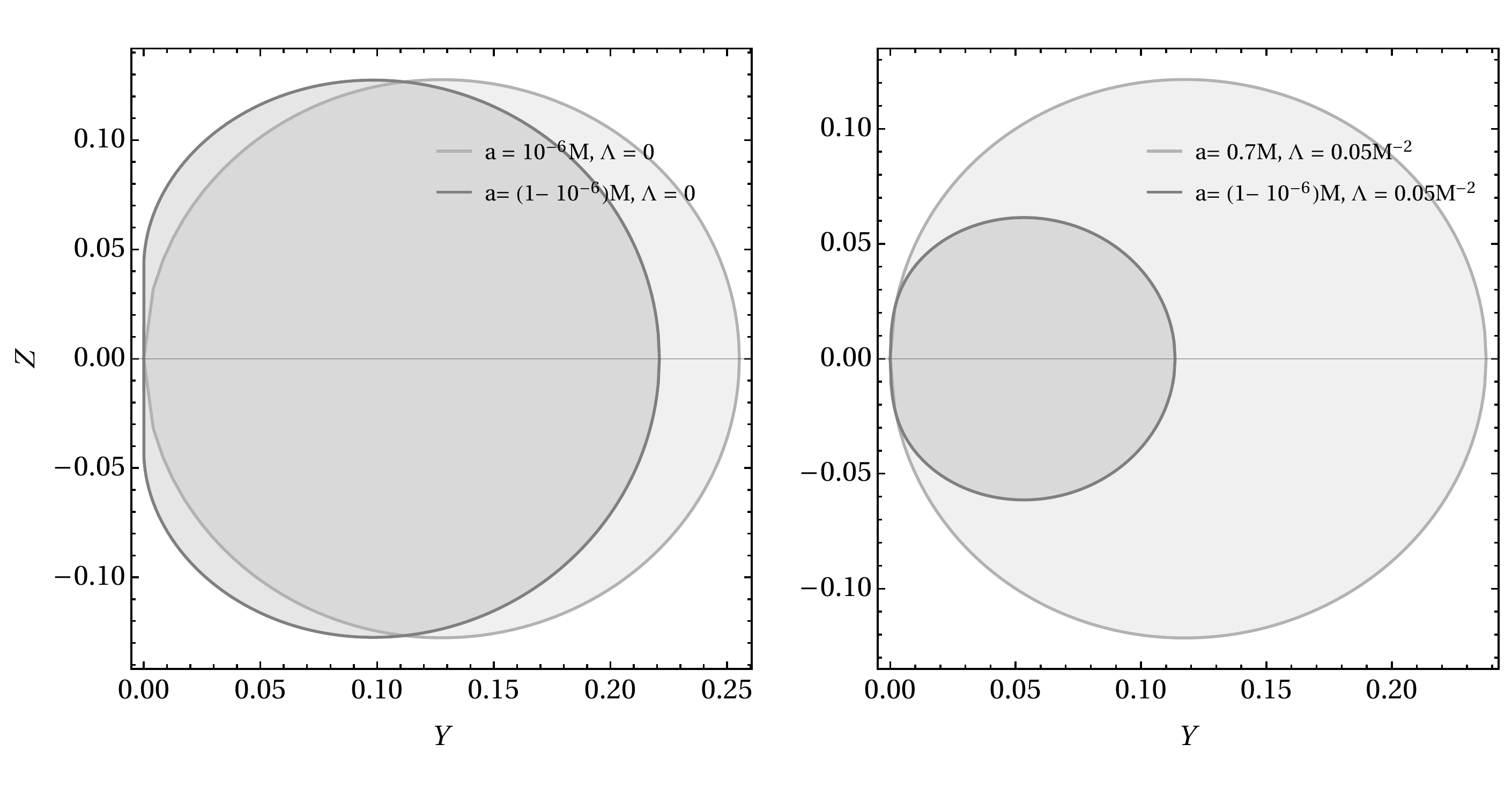}
	\caption{Shadow of Kerr-de Sitter black holes on projective plane $(Y,X)$ for distant observers. Left panel: Shadow of Kerr black holes for selected spin parameters for observers located at $r=40M$. Right panel: Shadow of Kerr-de Sitter black holes with cosmological constant $0.05M^{-2}$ for selected spin parameters for observers located cosmological horizon $r\approx 6.31M$.}
	\label{Fig8}
\end{figure}

\begin{figure}[ht]
	\centering
	\includegraphics[width=1\linewidth]{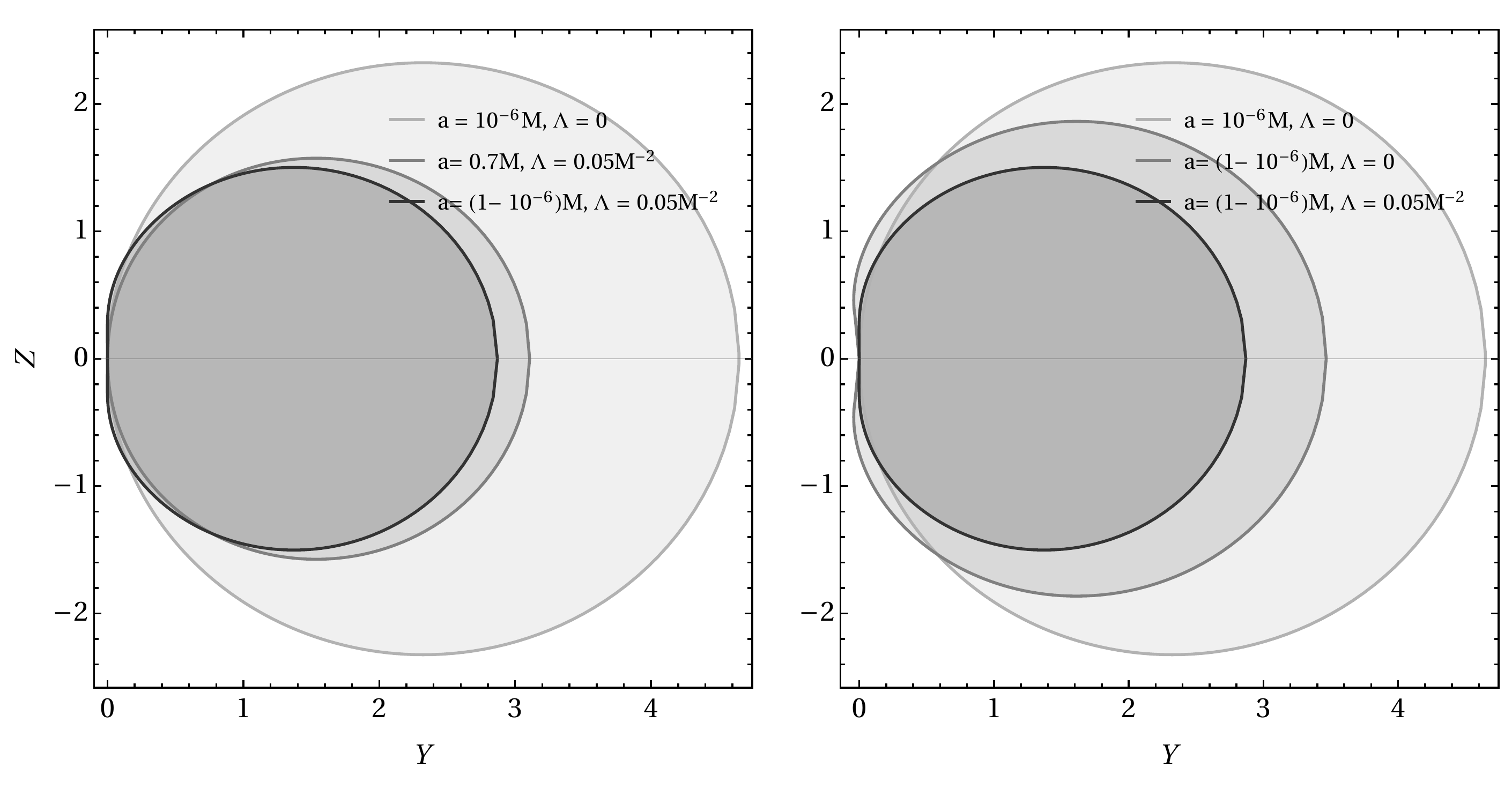}
	\caption{Shadow of Kerr-de Sitter black holes on projective plane $(Y,X)$ for observers located at $r=4M$. Left panel: Shadow for Kerr black holes for different spin parameters. Right panel: Shadow for Kerr-de Sitter black holes for different cosmological constant.  \label{FIg9}}

\end{figure}

In Figure~\ref{Fig8}, we plot shadow of Kerr(-de Sitter) black holes with selected
parameters for distant observers by using Eqs. (\ref{55}) and (\ref{56}). In the left panel of Figure~\ref{Fig8}, the shape of Kerr black hole shadow
seems the same as obtained by previous works \cite{bardeen_timelike_1973,grenzebach_photon_2014}. We would
 present further comparison in next section. As shown in the right panel of Figure~\ref{Fig8}, 
the shadow of Kerr-de Sitter black hole seems not serious distorted as spin parameter $a \rightarrow 1$. In Figure~\ref{FIg9}, we plot shadow of Kerr-de Sitter black holes with
selected parameters for observers closed to the black hole. It shows that the spin parameter
would cause distortion of shadow, while, cosmological constant could relieve
the distortion of the shadow. Besides, the shape of shadow for observers in
near region is different from the shape for distant observers. For example,
one can compare the shape of shadow for the Kerr black holes with $a = (1 - 10^{- 6}) M$ in Figures ~\ref{Fig8} and \ref{FIg9}.

\begin{figure}[ht]
	\centering
	\includegraphics[width=.7\linewidth]{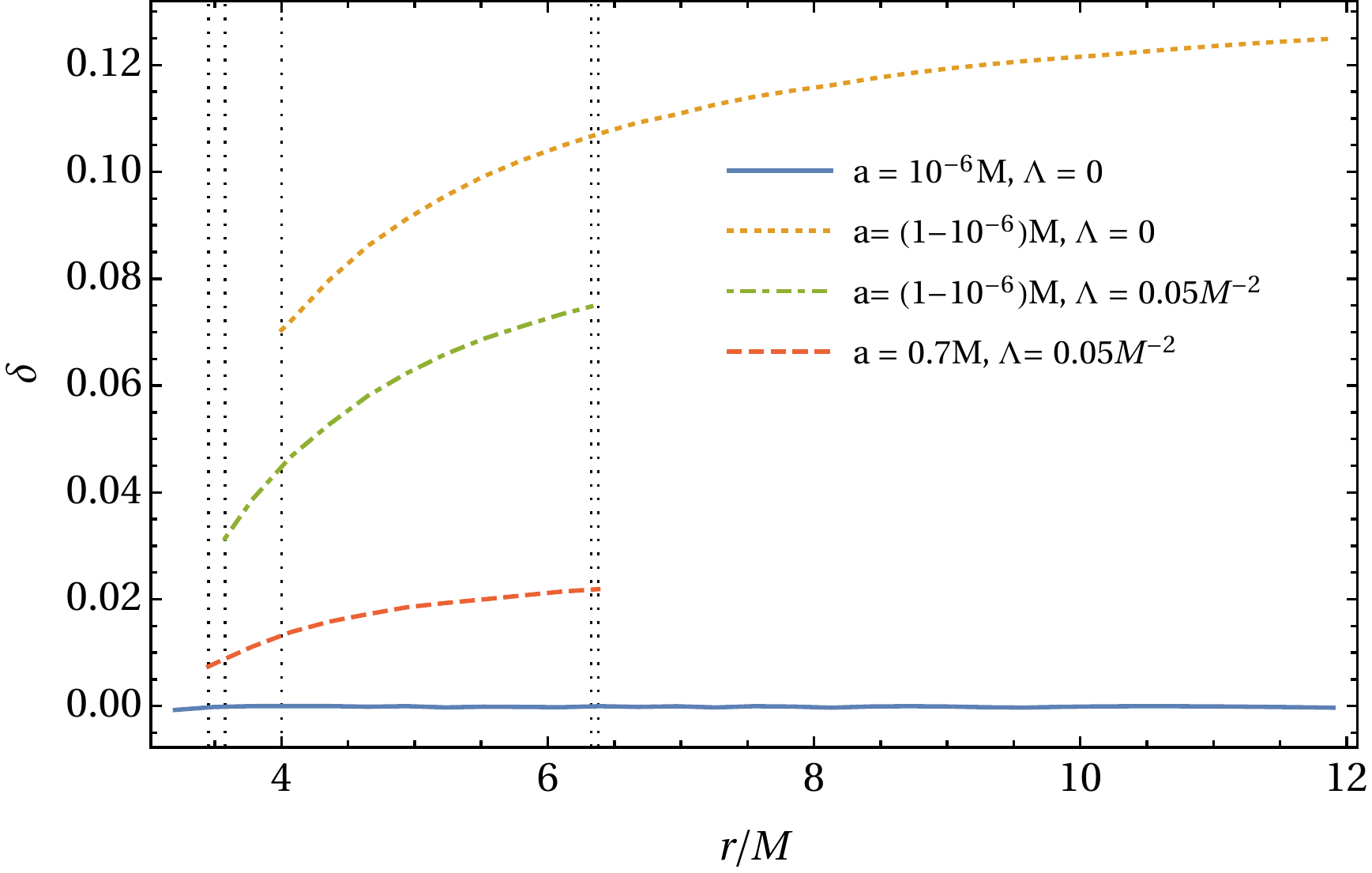}
	\caption{Distortion parameter as function	of distance with selected parameters.}
	\label{Fig10}
\end{figure}

For a  quantifiable description of the shape of shadow, we could introduce
 distortion parameter as
\begin{equation}
	\delta \equiv 1 - \frac{D_{\min}}{D_{\max}}~,
\end{equation}
where $D_{\min}$ and $D_{\max}$ are largest and smallest diameters of shadow,
respectively. This kind of quantity for black hole shadow was firstly proposed
by Hioki and Miyamoto \cite{hioki_hidden_2008}. In Figure~\ref{Fig10}, we plot distortion parameters as function
of distance with selected parameters. For Schwarzschild black holes, one can
deduce $\delta = 0$. The black hole with the largest
spin parameter and smallest cosmological constant has the most distorted shadow. For the rotating black holes,
the distortion parameters increases with the distance. It can be understood as part
of distortion of shadow attributed to accumulation of propagating
effect of light rays. For Kerr-de Sitter black holes, the static observers
located beyond cosmological horizon would not observe the shadow any more. And
there are not sudden changes of distortion parameter for these observers  near the cosmological horizon.

\begin{figure}[ht]
	\centering
	\includegraphics[width=1\linewidth]{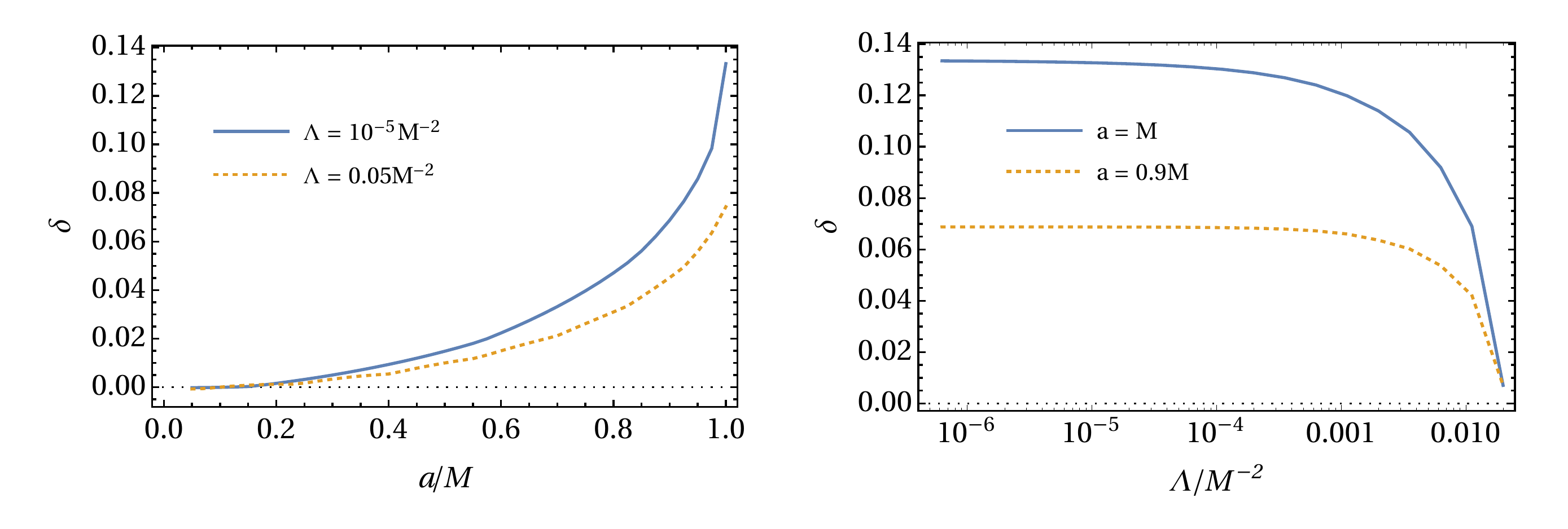}
	\caption{Left panel: Distortion parameters for observers on cosmological
		horizon as function of parameters $a$. Right panel: Distortion parameters for observers on cosmological
		horizon as function of cosmological constants $\Lambda$.}
	\label{Fig11}
\end{figure}

In Figure~\ref{Fig11}, we plot distortion parameters for observers on cosmological
horizon as function of parameters $a$ or $\Lambda$. The distortion parameters would
increase with spin parameter $a$ and decrease with cosmological constant
$\Lambda$. The cosmological constant less than $10^{- 4} M^{- 2}$ 
doesn't influence much on the distortion parameters of shadow. 

\section{Comparison with orthonormal tetrad approaches}\label{V}

We have shown the approach for calculating shadow of rotating black holes
without introducing orthonormal tetrad and applied it to Kerr-de Sitter black
holes as example. Last but not least, we should compare our results with
previous works. Here, we select two representative works of Bardeen \cite{bardeen_timelike_1973} and
Grenzebach et al. \cite{grenzebach_photon_2014}. Their approaches are both suited for studying the shadow with
respect to an observer located at finite distance.

\begin{figure}[ht]
	\centering
	\includegraphics[width=1\linewidth]{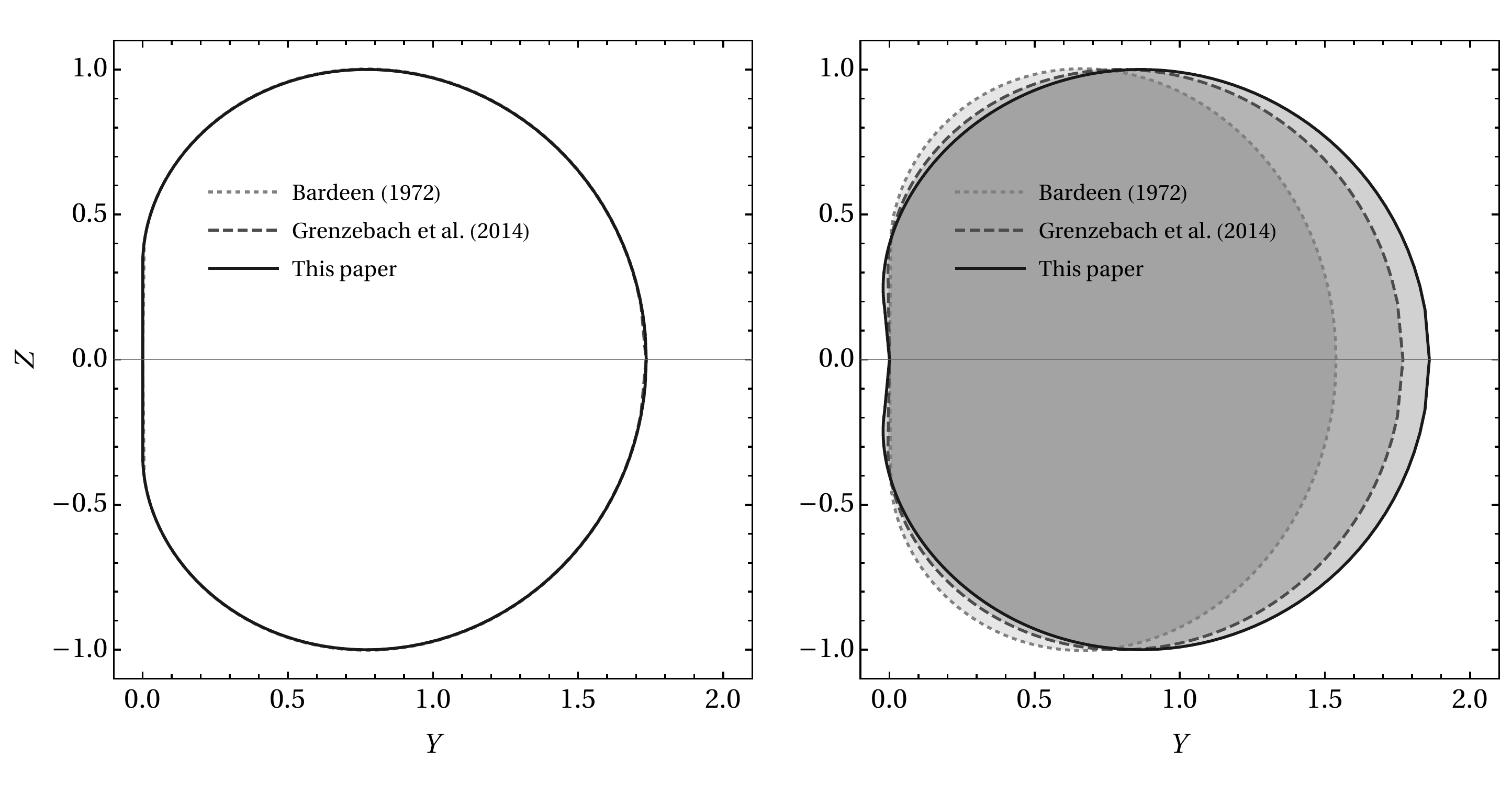}
	\caption{Scaling shadow of Kerr black holes in different approaches. Left panel: Shadow of Kerr black holes on  projective plane  $(X, Y)$ for observers located at $r=40M$. Right panel: shadow of Kerr black holes on  projective plane $(X, Y)$ for observers located at $r=4M$.}
	\label{Fig12}
\end{figure}

\begin{figure}[ht]
	\centering
	\includegraphics[width=0.7\linewidth]{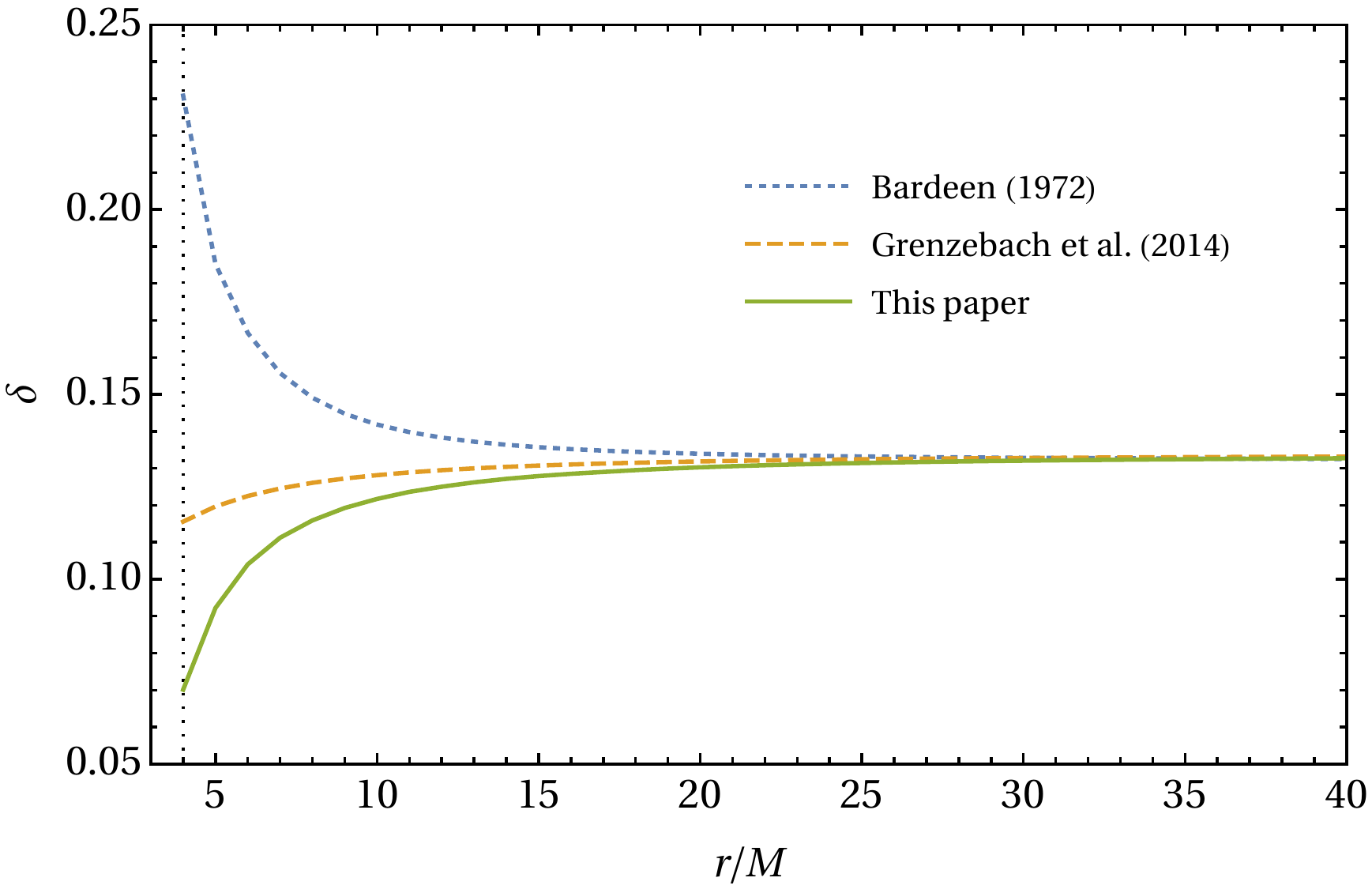}
	\caption{Distortion parameter as function of distance in different approaches for Kerr black holes.}
	\label{Fig13}
\end{figure}

In Figure~\ref{Fig12}, we present shadow of Kerr black holes on projective plane
$(X, Y)$ for observers in large distance and near region, respectively. For the sake of comparison, we use translation and scaling for the results in previous works. As shown in the left panel of Figure~\ref{Fig12}, the shape of Kerr
black hole shadow for distant observers in our approach is exactly the same as that in previous works
\cite{bardeen_timelike_1973,grenzebach_photon_2014}. In the right panel of Figure~\ref{Fig12}, Bardeen gave the most distorted shadow among others, while
we obtained a shadow with the smallest distortion. We plot distortion parameters as function of distance in these
different approaches for Kerr black holes in Figure~\ref{Fig13}. In large distance, the
distortion parameters  are exactly the same. In the near region, one
might find that Bardeen and Grenzebach et al. gave contrasty results of how the distortion parameter changes with distance. Our results are closed to that
obtained by Grenzebach et al.. Namely, the distortion parameter would increase with
distance. By the way, one might note that Bardeen's approach wasn't equipped
with stereographic projection. It suggests that his approach is
valid just in tendency for observers located at near region.

\begin{figure}[ht]
	\centering
	\includegraphics[width=1\linewidth]{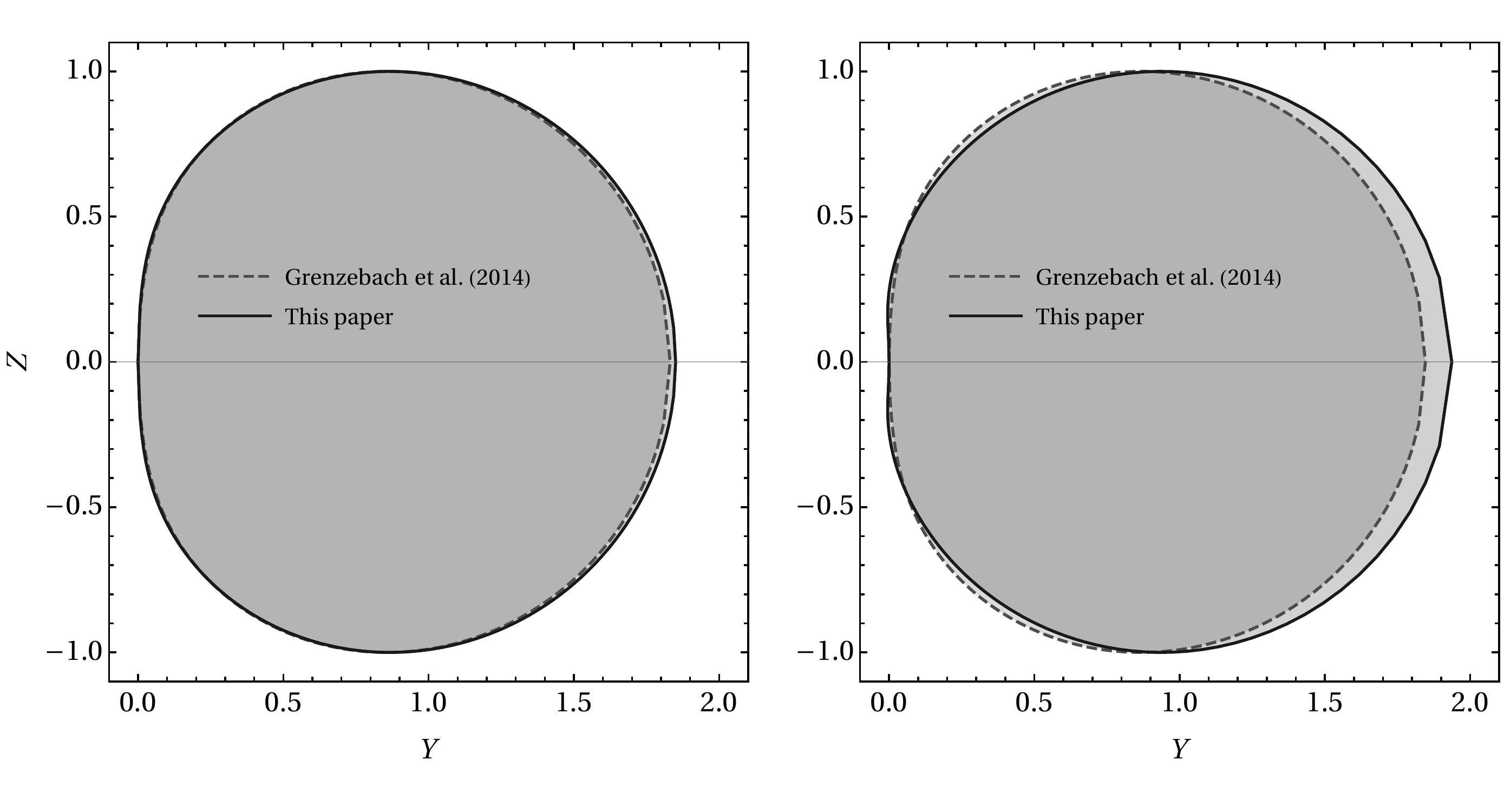}
	\caption{Shadow of Kerr-de Sitter black holes with cosmological constant $0.05M^{-2}$. Left panel: Shadow of Kerr-de Sitter black  holes on projective plane $(X, Y)$ for observers located  at cosmological horizon $r\approx 6.31M$. Right panel: Shadow of Kerr-de Sitter black  holes on projective plane $(X, Y)$ for observers located at outer boundary of photon region $r\approx 3.57M$}
	\label{Fig14}
\end{figure}

\begin{figure}[ht]
	\centering
	\includegraphics[width=0.7\linewidth]{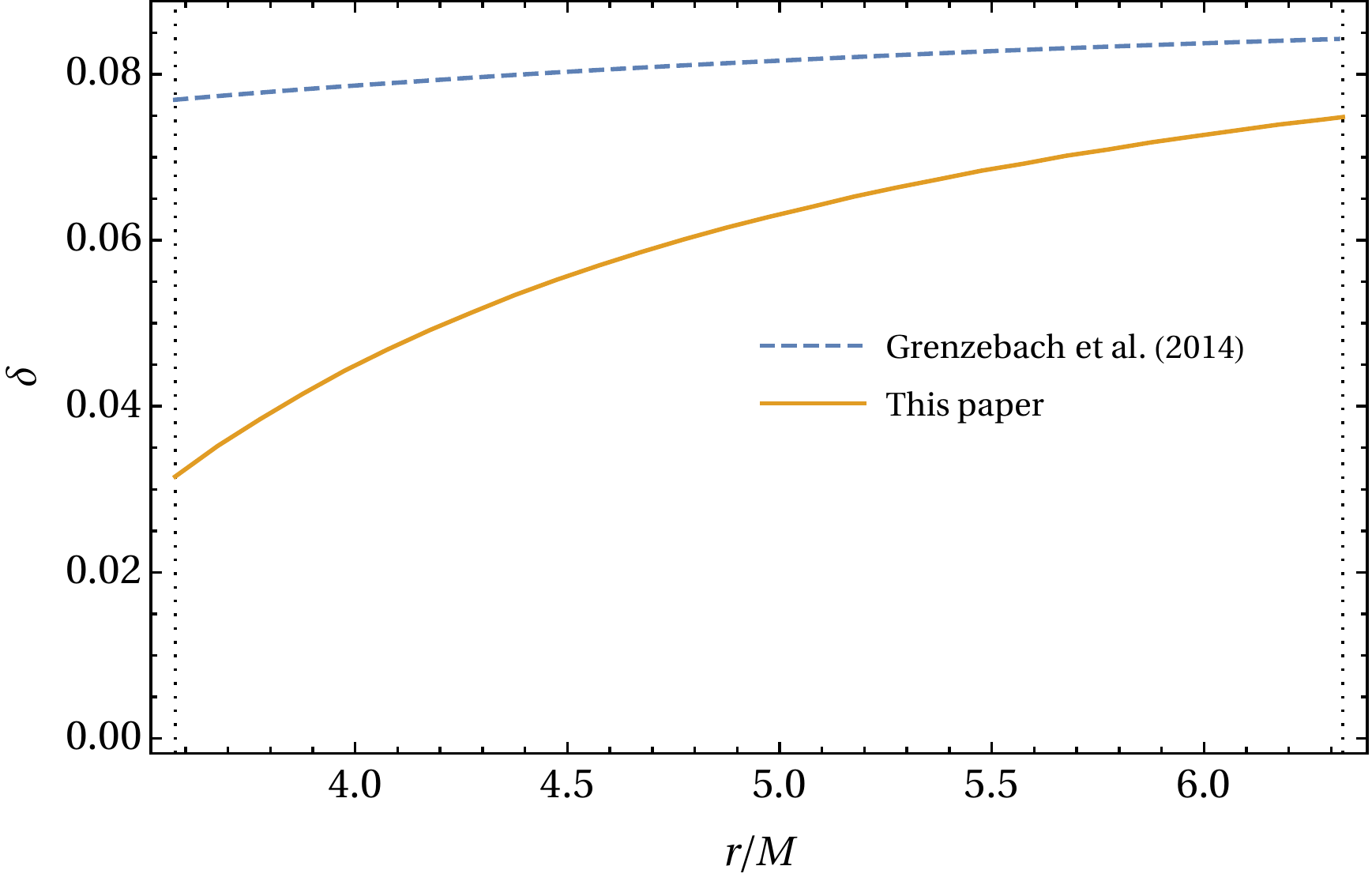}
	\caption{Distortion parameter as function of distance in different approaches for Kerr-de Sitter black holes.}
	\label{Fig15}
\end{figure}

For non-asymptotic space-time, we compare our results with
Grenzebach et al's \cite{grenzebach_photon_2014}. In Figure~\ref{Fig14}, we present shadow of Kerr-de Sitter black
holes on projective plane $(X, Y)$ for observers in large distance and near
region, respectively. The shadow obtained by Grenzebach et al is more distorted than
ours, especially for the observers in near region. In Figure~\ref{Fig15}, we also
compare the distortion parameters as function of distance in these approaches. The
distortion parameter of shadow in our approach is smaller and more sensitive to
distance. And differed from Kerr black holes, the distortion parameters in
different approaches can not be consistent with each other at certain
distance.

For those difference shown above, we think there might be two causes. Firstly, the
frame of ZAMOs used by Bardeen and Carter's frame used by Grenzebach et al., in fact,
suggest the reference frame adapted to different observers. It can be shown
via 0-component of these orthonormal tetrads in Kerr space-time \cite{bini_gyroscope_2017},
\begin{eqnarray}
	(e_0)_{\rm {ZAMO}} & = & \sqrt{\frac{(r^2 + a^2)^2 - a^2 \Delta \sin^2
			\theta}{\Delta \Sigma}} \left( \partial_0 + \frac{2 a   M
		r}{\sqrt{(r^2 + a^2)^2 - a^2 \Delta \sin^2 \theta}} \partial_3 \right) ~,
	\\
	(e_0)_{\rm {Carter}} & = & \frac{r^2 + a^2}{\sqrt{\Delta \Sigma}} \left(
	\partial_0 + \frac{a}{r^2 + a^2} \partial_3 \right) ~ .
\end{eqnarray}
And neither of them are adapted to a static observer, $u = \frac{1}{\sqrt{-
			g_{00}}} \partial_0$, which we used in Eqs.~(\ref{3}). In this sense, no one is
preferred than others in principle. It's just choice of different reference
frame. And all of them are deserved to be considered. Secondly, there is still
possibility that our approach with astrometric observables are fundamentally
different from the approaches with orthonormal tetrads. To confirm this possibility, further studies  are required. 

\section{Conclusions and discussions}\label{VI}

In this paper, we presented a new approach for calculating shadow of rotating
black holes with respect to observers at finite distance in formula of
astrometric observables. We obtained analytic formulas of  a general rotating black hole
shadow for given light rays from photon region. With the formulas, we studied size and shape of
Kerr-de Sitter black hole shadow as function of distance. In this space-time, we found shadow of Schwarzschild black holes  is the biggest and
shadow of Kerr black holes is most distorted. For distant observers, ours
results are consistent with previous works \cite{bardeen_timelike_1973,grenzebach_photon_2014}. For near-region observers, our
results are closed to Grenzebach et al.'s. Namely, the distortion parameters of shadow would
increase with distance.

Here, we only consider static observers fixed at inclination angle $\theta =
	0$ and $\theta = \frac{\pi}{2}$. In principle, it can be generalized into
arbitrary observers without any technical problems. Namely, one can substitute
4-velocity of observers $u$ in Eqs.~(\ref{4}) or (\ref{5}) by 4-velocity of arbitrary observers.

From studies on shadow of rotating black holes for observers located finite distance, one may get abundant
information involving the space-time geometry. Figure~\ref{Fig10} might suggest that part distortion of shadow can be
attributed to accumulation of propagating effect of light rays. 
Comparison between previous works with ours in section \ref{V} suggests
that motion status of observers is highly relevant to apparent shape of
rotating black holes, especially, for those observers located at finite
distance.

\acknowledgments{
	This work has been funded by the National Nature Science Foundation of China under grant No. 11675182 and 11690022.
}

\bibliography{citation}
\end{document}